\documentclass[twocolumn,showpacs,amsmath,amssymb,dvips,superscriptaddress,prl]{revtex4-1}
\usepackage{color,longtable}
\usepackage{graphicx}
\usepackage{amssymb,amsfonts,amsmath}
\usepackage{here}

\begin{document}
\title{Critical-like phenomenon in scraping of jamming systems}
\author{Masaya Endo and Rei Kurita}

\affiliation{%
Department of Physics, Tokyo Metropolitan University, 1-1 Minami-osawa, Hachiouji-shi, Tokyo 192-0397, Japan
}%
\date{\today}
\begin{abstract}
In jamming systems like colloids, emulsions, foams, and biological tissues, significant deformation is essential for processes such as material scraping or wound self-healing.
To adequately spread a foam or cream over a surface, external force must be applied to artificially scrape it.
The scraping of foam using a rigid plate has been observed to exhibit complex behavior distinct from that of simple liquids.
In this study, we quantitatively analyzed the transition between partial and slender scraping regimes by examining changes in internal structure and partial spreading lengths.
Our findings reveal that the sequential propagation of bubble rearrangement in the foam's internal structure leads to the partial scraping.
Moreover, the scraping length in the partial scraping regime shows divergence near the transition point, characterized by a critical exponent of approximately 0.61.
These results imply that foam scraping is governed by directional percolation theory, supported by the agreement between the experimentally observed critical exponent and theoretical predictions.
This research significantly advances the understanding of macroscopic kinetics and rheological behavior in jamming systems, including foams, colloids, emulsions, and biological tissues.
\end{abstract}

\keywords{Foam; Scraping; Rearrangement}

\maketitle

Jamming systems, including colloids, emulsions, foams, and biological tissues, are widely utilized in everyday applications such as paints, foods, heat insulation, and cellular processes.
For instance, foam exhibits exceptional heat insulation and mass transport shielding properties due to its composition, with approximately 90\% of its volume consisting of air.
This makes foam an effective fire-extinguishing agent to prevent oxidation, a cut off agent of oxygen supply by floating on foods or beverages and a widely used heat insulator, often placed between walls in construction~\cite{Weaire2001,cantat2013}.
Additionally, foam contributes to environmental sustainability by minimizing liquid waste and reducing material weight~\cite{Weaire2001, cantat2013}.
In applications such as coating surfaces with paints or foam or promoting biological wound healing, significant deformation of jamming systems is required, typically achieved by applying external forces~\cite{marchand2020, Deblais2015}.
However, unlike simple liquids, jamming systems generally cannot flow or spread easily due to their complex rheological properties~\cite{Merrer2012, Furuta2016, Katgert2013, siemens2010}.

The spreading of simple liquids over a substrate has been extensively studied, with the Landau-Levich-Derjaguin (LLD) theory providing the well-known equation $e \sim \kappa^{-1}(\frac{\eta U}{\gamma})^{2/3}$, where $\kappa^{-1}$ represents the capillary length, $\eta$ the liquid viscosity, $U$ the scraping velocity, and $\gamma$ the surface tension~\cite{taylor1962, deGennes2013, derjaguin1941, levich1942}.
The LLD theory, derived from the Navier-Stokes equation while neglecting inertia, aligns well with experimental observations and has been extensively utilized to predict film thickness, particularly in dip-coating processes.
Recent advancements in research on scraping phenomena include studies of scraping using elastic scrapers for simple liquids~\cite{seiwert2013, krapez2020}.
In contrast, the rheological properties of jamming systems exhibit shear-thinning behavior, markedly different from simple liquids~\cite{kraynik1988, cohen2013, hohler2005, Deblais2015}.

Recent studies have revealed that the scraping of foams exhibits intricate behaviors~\cite{Endo2023}.
On hydrophobic substrates, foam can be scraped uniformly at low scraping velocities, maintaining the same width and a thickness equal to the gap height between the plate and the substrate (homogeneous scraping).
At medium scraping velocities, the foam partially slips, resulting in incomplete scraping (partial scraping).
In the high-velocity regime, the foam is scraped again, but with a narrower width compared to the initial foam layer (slender scraping).
Previous research demonstrated that at low scraping velocities, bubbles anchor to the substrate, enabling homogeneous scraping~\cite{Endo2023}.
However, the mechanism underlying the scraping behavior at high velocities remains unclear.

To address this, we conducted a quantitative investigation into the transition mechanism from partial scraping to slender scraping, analyzing both macroscopic characteristics and localized internal bubble rearrangements.
Our findings indicate that this transition is driven by a critical-like phenomenon involving sequential bubble rearrangements under large deformation in jamming systems.
These insights significantly advance the understanding of macroscopic kinetics and rheological behavior in jamming systems.

Here, we briefly outline the physical properties of the foams and the methods used in this study. Comprehensive experimental procedures are detailed in the Supplementary Information~\cite{supple}.
Foams were prepared using a 5.0 wt\% solution of the ionic surfactant TTAB (tetradecyl trimethyl ammonium bromide) dissolved in glycerol and deionized water.
This concentration exceeds the critical micelle concentration (0.12 wt\%)~\cite{Danov2014}, and a 3.0 wt\% solution exhibits a surface tension of 37 mN/m.
The interface rigidity of TTAB is known to be relatively low~\cite{Yanagisawa2023}.
Foams were generated with a foam dispenser (Daiso Ind. Co., Ltd., Japan), resulting in a liquid fraction of $\phi = 0.10 \pm 0.01$.
The mean bubble diameter $d$ was 0.26 mm, with a standard deviation of 0.07 mm.

A hard elastic sheet made of polydimethyl-siloxane (Correcsil puls, Yamahachi Dental MFG., Co., Japan) was primarily used as a partially wetting substrate.
An acrylic plate was positioned perpendicular to the substrate with a gap $b$, which was varied between $b = 1.0 - 2.5$ mm.
The confinement length $L$ was independently adjusted in the range of $L = 1 - 10$ mm.
For the foam scraping diagram, a hemispherical foam with a diameter $W$ was placed on the substrate.
The foam was spread by moving the substrate horizontally at a constant velocity of $V = 1.0 - 35.0$ mm/s using an electric slider (LTS300/M, Thorlabs, Inc., US).
The scraping dynamics were captured from above using a video camera (EOS R, Canon Inc., Japan).
The experimental setup for observing internal bubble rearrangements in the foam's cross-section is illustrated in Fig. S1~\cite{supple}.

\begin{figure}[htbp]
\begin{center}
\includegraphics[width=8cm]{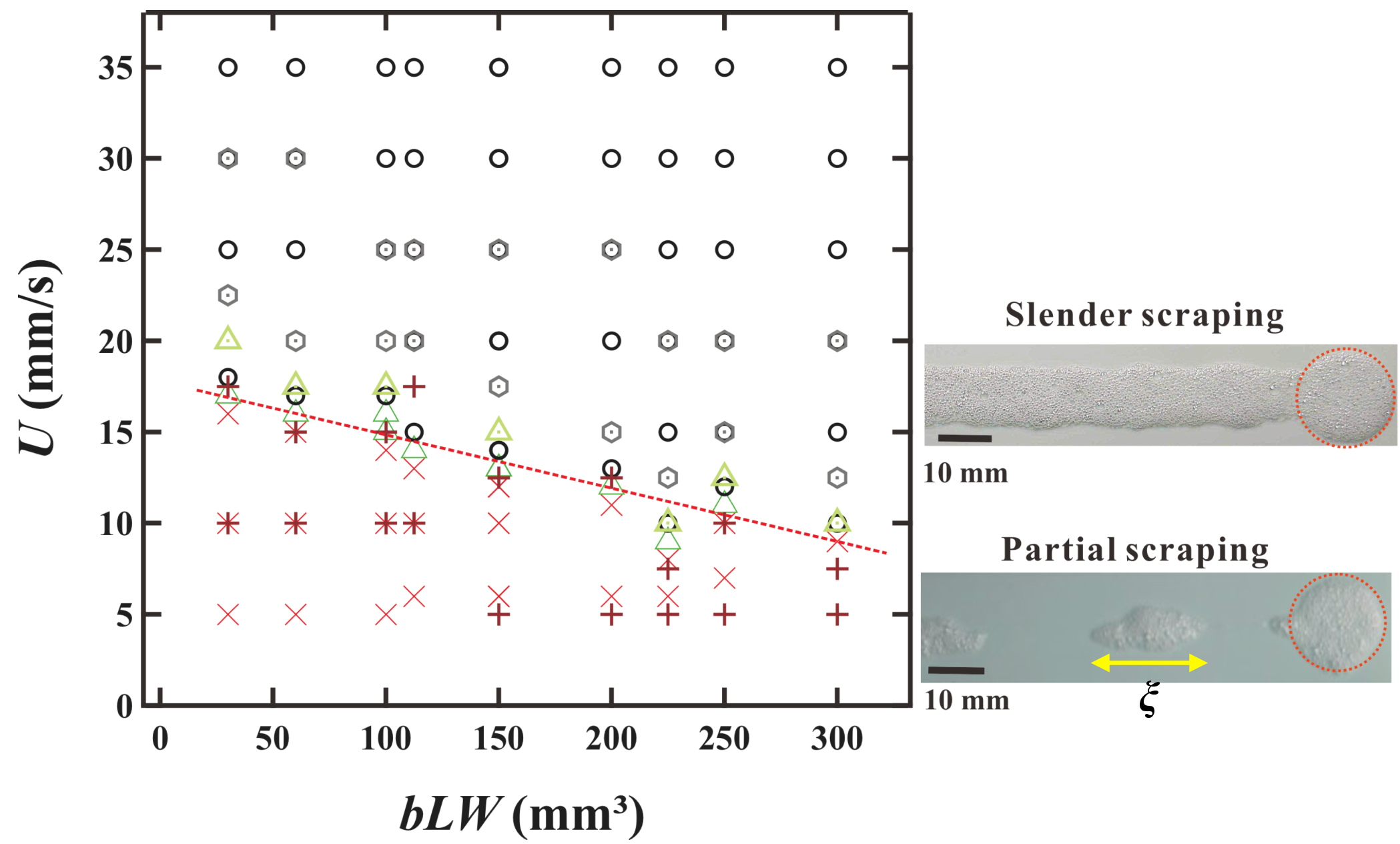}
\caption{Diagram of scraping patterns showing the relationship between $U$ and $bLW$.
The transition boundary between the partial scraping (cross symbols) and the slender scraping (open symbols) is influenced by the volume beneath the plate.
For instance, for parameter sets $(b, L, W)$ = (1.5 mm, 5.0 mm, 20 mm), (1.0 mm, 5.0 mm, 30 mm), and (2.0 mm, 3.0 mm, 25 mm), the critical velocity $U_c$ remains consistent at 14 mm/s.
It is important to note that the dependencies of $U_c$ on $W$ and $b$ cannot be accounted for solely by the macroscopic shear rate $U/b$.
}
\label{Diagram}
\end{center}
\end{figure}

Initially, we examined the transition boundary between partial scraping and slender scraping by varying the scraping velocity $U$, the gap beneath the plate $b$, the plate width $L$, and the foam width $W$.
The boundary separating partial scraping and slender scraping shifted downward with an increase in any of the parameters $b$, $L$, or $W$, as depicted in Fig. S2~\cite{supple}.
The phase diagram showing the relationship between the velocity $U$ and the volume beneath the plate $bLW$ is presented in Fig.~\ref{Diagram}.
For different parameter combinations, such as $(b, L, W)$ = (1.5 mm, 5.0 mm, 20 mm), (1.0 mm, 5.0 mm, 30 mm), and (2.0 mm, 3.0 mm, 25 mm), the boundary velocity consistently converged to approximately $U_c$ = 14 mm/s across all cases.
These findings suggest that the critical velocity $U_c$, marking the transition from partial scraping to slender scraping, is governed by the volume beneath the plate.

In this context, a yield stress is typically associated with the fluidization of foam, as seen in slender scraping.
The shear stress acting on the bubbles can be estimated as $P = \eta U_c / b \sim 0.7$ Pa, whereas the yield stress is approximately 20 Pa~\cite{Weaire2008}.
Therefore, the transition occurs under a stress much lower than the yield stress.
Moreover, the boundary line shows a steeper decline as a function of $U_c/b$.
Furthermore, $U_c$ also depends on $W$, which is oriented perpendicular to the shear direction.
Consequently, the dynamic transition cannot be explained solely by shear stress.

\begin{figure}[htbp]
\begin{center}
\includegraphics[width=8cm]{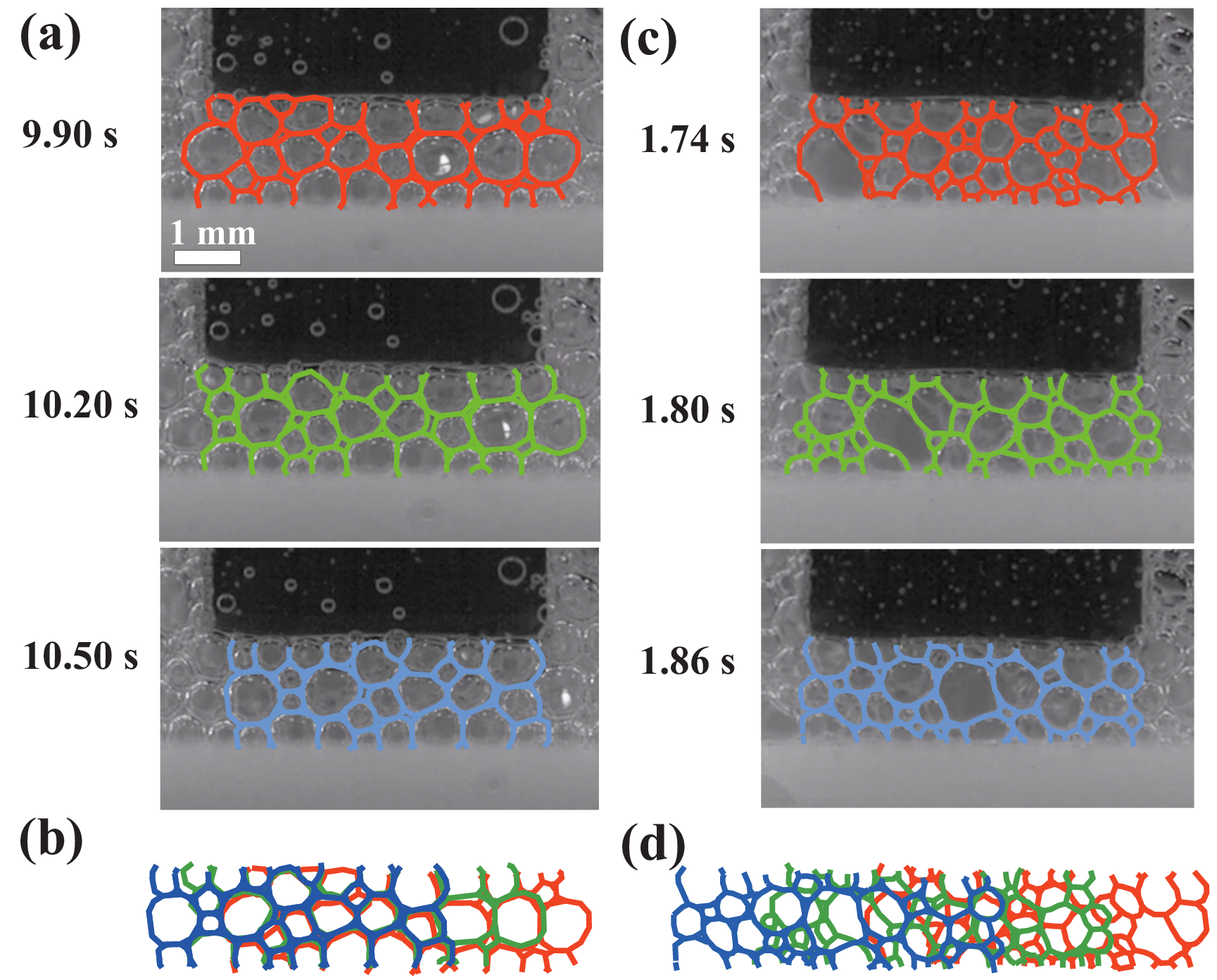}
\caption{
(a) The extracted interfaces of the bubbles at various times in the homogeneous scraping.
(b) A superimposed figure after applying the transformation $x^{\prime} = x - Ut$ to each interface in (a). In the homogeneous scraping, the bubbles move horizontally along with the substrate without undergoing any rearrangement, which is consistent with the foam anchoring to the substrate in this pattern.
(c) The extracted interfaces of the bubbles at different times in the slender scraping.
(d) A superimposed figure after applying the transformation $x^{\prime} = x - Ut$ to each interface in (c). In contrast to the homogeneous scraping, the slender scraping shows that the bubbles have rearranged over time.
}
\label{rearrangement}
\end{center}
\end{figure}

Next, we examined the internal structure of the bubbles during scraping. 
Figure \ref{rearrangement}(a) presents the extracted interfaces of the bubbles at various times in the homogeneous scraping. 
Figure \ref{rearrangement}(b) shows the superimposed interfaces after the transformation $x^{\prime} = x - Ut$, where the interface coordinates are adjusted based on the distance the substrate has moved. 
In the homogeneous scraping, the interfaces align perfectly at all times, indicating that the bubbles are moving in parallel with the substrate. This observation is consistent with the behavior where the internal bubbles at the bottom are anchored to the substrate in the homogeneous scraping~\cite{Endo2023}. 
Similarly, the behavior of the internal bubbles in the slender scraping was analyzed. 
Figure \ref{rearrangement}(c) displays the extracted bubble interfaces at different times in the slender scraping, and panel (d) presents the superimposed interfaces after applying the transformation $x^{\prime} = x - Ut$. In the slender scraping, the interfaces of the bubbles beneath the plate do not overlap, suggesting that the bubbles undergo rearrangement over time.

\begin{figure}[htbp]
\begin{center}
\includegraphics[width=8cm]{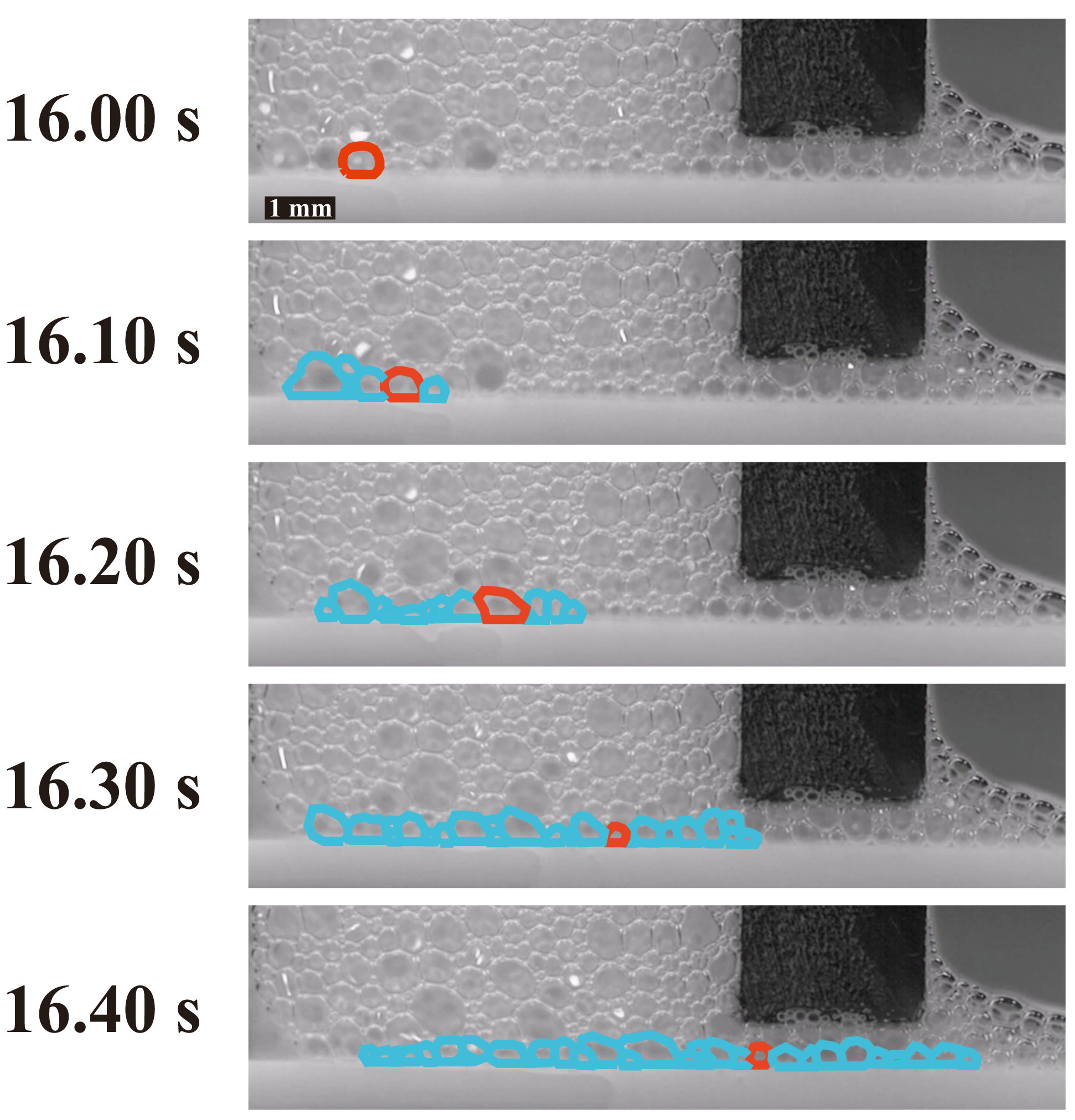}
\caption{The time evolution of bubble positions during partial scraping at $U$ = 10.0 mm/s and $b$ = 1.5 mm. A bubble, highlighted with a red surface, undergoes the first rearrangement, followed by the sequential rearrangement of the surrounding bubbles, indicated in blue. }
\label{rearrangement2}
\end{center}
\end{figure}

We also examined the changes in the internal structure of the foam during partial scraping. In the partial scraping regime, the bubbles typically slide on the substrate for most of the time, and occasionally, a rearrangement of the bubbles occurs beneath the plate, leading to partial scraping. In rare instances, a rearrangement event takes place far from the plate, as shown in Fig.~\ref{rearrangement2}.
Figure \ref{rearrangement2} illustrates the time evolution of bubble positions at $U$ = 10.0 mm/s and $b$ = 1.5 mm. Just before partial scraping begins, a bubble, highlighted in red, undergoes a rearrangement, followed sequentially by the rearrangement of surrounding bubbles (blue bubbles). This propagation of rearrangement events eventually passes beneath the plate, leading to partial scraping  (see Supplementary movie.S1). To further investigate the role of this sequential rearrangement, we intentionally induced a rearrangement within the partial scraping regime. By inserting a hydrophobic elastic sheet into the foam, the internal bubbles were rearranged as they were pulled by the sheet movement (see Supplementary movie.S2)~\cite{marchand2020, Endo2023, supple}. 
Upon insertion and removal of the elastic sheet, the partial scraping occurs as shown in Fig.~S3~\cite{supple}.

When the foam slips on the substrate, the drag force exerted by the substrate is balanced by viscous dissipation. However, this balance is locally disrupted by the flow generated by the bubble rearrangement. The disruption of this balance propagates through the surrounding area. We also observe that the probability of sequential rearrangement is influenced by $U$. When $U$ is small, the sequential rearrangement halts midway, resulting in the partial scraping. In contrast, for $U \ge U_c$, the sequential rearrangement continues, leading to the slender scraping.

\begin{figure}[htbp]
\begin{center}
\includegraphics[width=8cm]{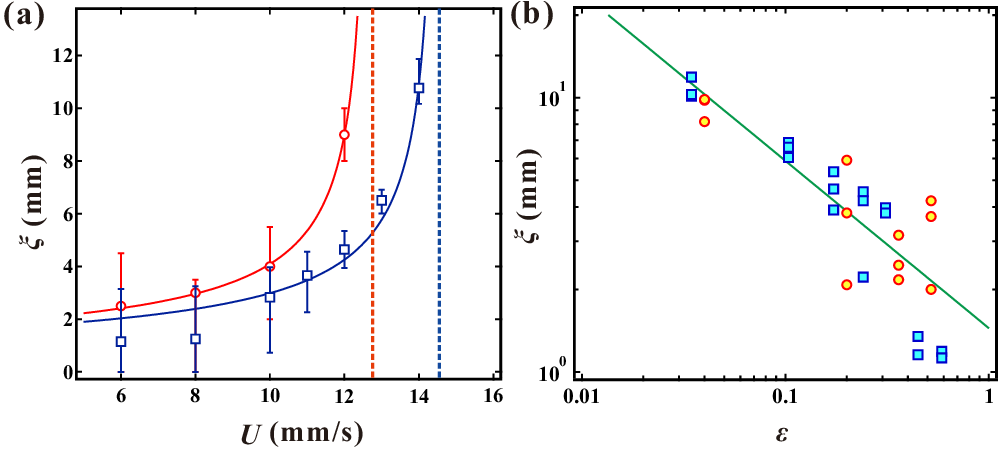}
\caption{(a) The partial scraping foam length, $\xi$, is plotted as a function of $U$. The square and circle symbols correspond to $\xi$ values at $b$ = 1.0 mm and $b$ = 1.5 mm, respectively. The symbols represent the averages from three measurements, with the error bars indicating the range between the maximum and minimum values.
(b) $\xi$ is shown as a function of $\epsilon$ where $\epsilon = (U_c - U)/U_c$. The data reveals that $\xi \sim \epsilon^{-\alpha}$, with $\alpha = 0.61 \pm 0.06$. This exponent, $\alpha$, is consistent with the prediction from the directional percolation theory.
}
\label{critical}
\end{center}
\end{figure}

Finally, we investigated the scraping length $\xi$ in the scraping direction as a characteristic of the partial scraping regime.
Figure \ref{critical}(a) shows the dependence of $\xi$ on $U$ for $b$ = 1.0 mm (square) and $b$ = 1.5 mm (circle).
The symbols represent the averages from three measurements, with error bars corresponding to the maximum and minimum values.
We observed that $\xi$ increases sharply as the transition rate $U_c$ is approached.
In addition, the dimensionless reduced velocity $\epsilon = (U_c - U) / U_c$ is introduced, and its dependence on $\xi$ is presented in Fig.~\ref{critical}(b).
The $\xi$ values for $b$ = 1.0 mm and 1.5 mm can be scaled by $\epsilon^{-\alpha}$ within the error range, resulting in $\alpha = 0.61 \pm 0.06$.

Those sequential events have been also observed in phenomena such as the transition from laminar to turbulent flow in fluids. 
In the laminar-turbulent transition, the shift to turbulence occurs at a much lower Reynolds number than predicted by theory, and the transition to turbulence propagates sequentially when part of the laminar flow is disturbed~\cite{Avila2011, Sano2016}. 
This sequential propagation of turbulence can be explained by the critical phenomena associated with directional percolation (DP) theory~\cite{Sano2016, Haye2000, Takeuchi2009}. 
This behavior is similar to the sequential rearrangement of bubbles in the foam scraping.
Moreover, the critical exponent of 0.61 observed in foam scraping is consistent with the theoretical DP exponent of 0.58 in three dimensions~\cite{Haye2000}. This suggests that the transition observed in foam scraping can be understood through the framework of directional percolation theory.


Finally, it has been reported that sequentiality is influenced by the liquid fraction~\cite{Yanagisawa2023}, and the rearrangement behavior is also dependent on the dispersion of bubble size~\cite{Yanagisawa2021a}. In this study, however, the liquid fraction and bubble size dispersion are nearly constant. Future research will explore the dependence of $U_c$ on liquid fraction and bubble size dispersion to better understand the factors that determine $U_c$.

In summary, jamming systems are widely utilized in various fields such as heat retention, construction, food, and transportation by being applied to surfaces~\cite{Weaire2001, cantat2013}. However, due to their complex rheology, jamming materials do not spontaneously spread on surfaces like simple liquids. As a result, an external force is necessary to scrape them onto surfaces artificially. Recent studies have shown that the scraping behavior of foam exhibits complex dynamics~\cite{Endo2023}. At low scraping velocities, foam undergoes homogeneous scraping as a result of anchoring to the substrate~\cite{Endo2023}. At intermediate velocities, the foam experiences partial scraping, while at high velocities, slender scraping occurs. However, the nature and mechanism of this transition remain unclear.

In this study, we quantitatively examined the transition from partial scraping to slender scraping. It was found that foam undergoes fluidization at stress levels much lower than the yield stress, and the transition boundary is governed by volume rather than gap size, which contradicts the shear stress concept. Our findings reveal that a single event of bubble rearrangement at a specific location propagates sequentially to surrounding areas, leading to the partial spreading. Moreover, the scraping distance in the partial scraping regime exhibits divergent behavior as it approaches the transition point, with a critical exponent of approximately 0.61. These results suggest that the dynamical transition can be described by directional percolation theory~\cite{Sano2016, Haye2000, Takeuchi2009}.

The microscopic sequential rearrangement of bubbles in the foam is likely to be a universal phenomenon in other jamming systems, such as colloidal suspensions, emulsions, and biological systems like cells. Therefore, this study contributes significantly to the understanding of macroscopic large deformations and the rheology of jamming systems.

R. K. was supported by JSPS KAKENHI Grant Number 20H01874.

R.~K. conceived the project. M.~E. performed the experiments and analyzed the data. 
R.~K. wrote the manuscript.

The authors declare that they have no competing interests. 

Correspondence and requests for materials should be addressed to R.~K. (kurita@tmu.ac.jp).

All data generated or analyzed during this study are included in this published article.

%

\end{document}